# The Phase Structure of Metallic Money: An MPTT Framework for the Spanish Price Revolution

**Running title:** Two-phase metallic-money transmission in early modern Spain

Ran Huang

Academy for Small Commodity Economy, Yiwu Research Institute of Fudan University, Yiwu 32200, China

huangran@fudan.edu.cn

## Abstract

The Spanish Price Revolution is usually treated as a classical case in which American bullion inflows expanded the money supply and generated inflation. This view captures the first phase of the episode but fails to explain why the same monetary expansion did not continue to produce proportional price growth after 1600. We develop a two-phase Money Phase Transition Theory (MPTT) model in which the classical monetary relation is recovered before a transition point, while a second-phase correction term modifies the money-price transmission coefficient after the transition. Using annual Spanish CPI and reconstructed money-supply data, we show that 1500–1600 was a high-transmission metallic inflationary phase: CPI increased approximately 3.35-fold while money supply increased approximately 3.73-fold. After 1600, money supply continued to rise, increasing approximately 1.82-fold during 1600–1650, while CPI rose only approximately 1.22-fold. A classical one-phase model fitted on 1500–1600 therefore overpredicts post-1600 prices when extrapolated forward. The MPTT two-phase model with transition point $\tau = 1600$ estimates $\beta_1 \approx 0.949$, $\gamma \approx -0.812$, and $\beta_2 = \beta_1 + \gamma \approx 0.137$, indicating a sharp post-transition weakening of monetary transmission. An unrestricted break scan identifies a deeper BIC-minimizing break around 1636. These results suggest that the Spanish Price Revolution was not a single monotonic bullion-inflation process, but the rise and exhaustion of high-transmission metallic-money inflation.

## 1. Introduction

The Spanish Price Revolution has long served as one of the canonical historical examples of monetary expansion producing inflation. From the sixteenth century onward, large quantities of American silver and gold entered the Spanish and European monetary system, while prices rose persistently over the early modern period (Hamilton, 1934; Fisher, 1911). Recent reconstruction of the Spanish money supply confirms the broad monetary character of the episode: the Spanish money stock, measured in silver-equivalent terms, increased more than tenfold between 1492 and 1810, and this expansion can account for much of the long-run rise in the price level from the perspective of the equation of exchange (Chen, Palma, and Ward, 2021). The Spanish case is therefore not one in which money was irrelevant. It is a case in which the monetary interpretation is almost unavoidable.

Yet the deeper question is not whether money mattered, but whether the money-price transmission relation remained stable. A simple bullion-inflation narrative suggests a monotonic mechanism: more precious metal entered the monetary system, money became more abundant, and prices rose. This captures the first-order fact, but it does not fully describe the internal structure of the episode. During 1500–1600, money and prices rose broadly together. After 1600, however, money supply continued to expand while CPI growth flattened. The classical monetary relation therefore works in the first phase but fails by extrapolation after the transmission break.

This paper approaches the Spanish Price Revolution through Money Phase Transition Theory (MPTT). The key claim of MPTT is that monetary quantity affects prices through a phase-dependent transmission coefficient. The theory does not deny the classical monetary relation. Instead, it treats the classical relation as a one-phase approximation. When the monetary system enters a different functional phase, the transmission coefficient can change.

The simplest empirical form of MPTT is a two-phase correction to the classical log-level monetary relation:

$$\ln P_t = a + \beta_1 \ln M_t + \gamma H(t-\tau)(\ln M_t - \ln M_\tau) + \varepsilon_t,$$

where $P_t$ is the price level, $M_t$ is money supply, $\tau$ is the transition point, $H(t-\tau)$ is a step function that activates after the transition, and $\gamma$ is the phase-correction parameter. Before the transition, the model reduces to the classical relation. After the transition, the effective transmission coefficient becomes

$$\beta_2 = \beta_1 + \gamma.$$

Thus, $\gamma$ measures the change in monetary transmissibility. If $\gamma = 0$, the classical one-phase model is sufficient. If $\gamma < 0$, the second phase weakens money-price transmission. If $\gamma > 0$, the second phase amplifies it.

This formulation clarifies the relation between MPTT and earlier monetary-function theory. The earlier SCR framework addressed the functional logic of currency, reservation, and nominal-value preservation (Huang, 2018). MPTT is a distinct but compatible theory: it does not primarily claim that money must be divided into fixed circulation and reserve compartments. Rather, it asks whether the empirical transmission coefficient from money to prices changes across monetary phases. In this sense, MPTT is not a renaming of SCR, but a new phase-transmission theory built on a different empirical object.

The Spanish case provides a clean historical test (Fig. 1). If the classical relation is fitted on 1500–1600 and then extrapolated to 1700, it overpredicts the post-1600 price level. The gap between observed prices and classical extrapolation is precisely what the MPTT correction term is designed to capture. In Spain, the estimated phase correction is strongly negative. With $\tau = 1600$, the pre-transition coefficient is close to unity, while the post-transition coefficient falls to approximately 0.14. This means that the same kind of monetary expansion that was highly inflationary in the sixteenth century became weakly inflationary after the transition.

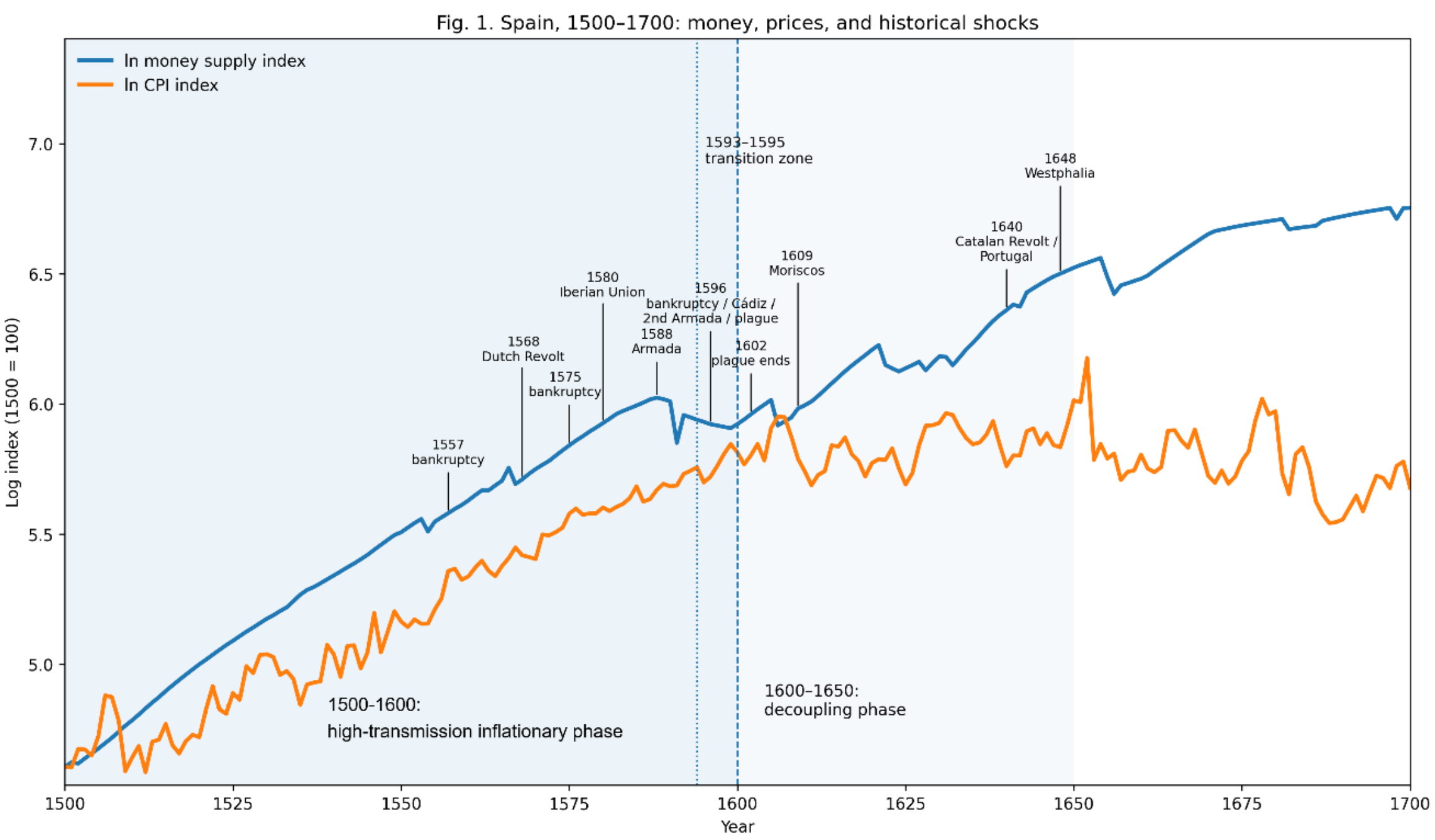

**Fig. 1. Two-regime structure of Spanish metallic-money transmission, 1500–1700.** Log-indexed trajectories of Spanish money supply and CPI, normalized to 1500 = 100, with major fiscal, military, demographic, and social shocks overlaid on the same historical axis. The figure separates the pre-1600 high-transmission inflationary phase, during which money and prices rose broadly together, from the post-1600 decoupling phase, during which money supply continued to expand while CPI growth flattened. Historical markers include the Habsburg bankruptcies, the Dutch Revolt, the 1588 Armada defeat, the 1596 fiscal and naval crisis cluster, the 1596–1602 plague wave, and the 1609 expulsion of the Moriscos. CPI data are based on Álvarez-Nogal and Prados de la Escosura (2013); money-supply data are based on Chen, Palma, and Ward (2021).

The timing is also historically plausible. The late sixteenth and early seventeenth centuries were not a tranquil continuation of earlier prosperity. Philip II's monarchy faced repeated fiscal crises and sovereign defaults, including major defaults in 1557, 1575, and 1596 (Drelichman and Voth, 2011, 2014). The Dutch Revolt began in 1568 and imposed a long fiscal-military burden. The Spanish Armada failed in 1588 (Parker, 1998). The 1590s also brought the 1596 fiscal crisis, the Cádiz shock, a second Armada disaster, and the plague wave of 1596–1602. These events do not by themselves prove the monetary transition, but they show that the post-1600 weakening of transmission occurred in an environment of imperial overstretch, fiscal absorption, demographic disruption, and possible circulation breakdown.

The contribution of this paper is therefore twofold. First, it revises the interpretation of the Spanish Price Revolution. Spain was not merely a monotonic bullion-inflation episode; it contained a high-transmission metallic inflationary phase followed by a weakened-transmission phase. Second, it proposes a simple and portable MPTT empirical equation. The classical monetary relation is the first-phase limit, and the phase correction parameter $\gamma$ measures whether the second phase amplifies or exhausts monetary transmission. In the Spanish case, $\gamma < 0$, indicating the exhaustion of high-transmission metallic-money inflation after 1600.

## 2. Data and historical setting

The empirical analysis uses an annual Spanish panel combining a historical consumer price index and reconstructed money-supply estimates. The CPI series is based on Álvarez-Nogal and Prados de la Escosura (2013). The money-supply series is based on Chen, Palma, and Ward (2021). The full assembled panel spans 1492–1810, but the main analysis focuses on 1500–1700 in order to capture both the sixteenth-century high-transmission phase and the post-1600 weakening of transmission.

Inflation and money growth are computed as

$$\pi_t = 100\Delta \ln P_t,$$

and

$$\mu_t = 100\Delta \ln M_t,$$

where $P_t$ is the CPI and $M_t$ is annual money supply. However, the central object of the revised model is not the annual growth-rate correlation. It is the log-level transmission coefficient

$$\beta = \frac{d\ln P}{d\ln M}.$$

Figure 1 provides the basic empirical setting. It plots log-indexed money supply and CPI from 1500 to 1700 and overlays major fiscal, military, demographic, and social shocks surrounding the transmission break. Table 1 summarizes the two-regime pattern quantitatively. During 1500–1600, CPI increased approximately 3.35-fold while money supply increased approximately 3.73-fold, yielding a log-level transmission ratio of approximately 0.92. During 1600–1650, money supply continued to increase by approximately 1.82-fold while CPI increased only approximately 1.22-fold, reducing the log-level transmission ratio to approximately 0.33.

This pattern is the empirical basis for the MPTT two-phase model. The question is not whether money and prices rose together in the long run. They did. The question is why the first-phase relation failed after the transition.

## 3. Theory and empirical model: the MPTT two-phase transmission equation

The empirical pattern in Fig. 1 suggests that the Spanish Price Revolution cannot be represented as a single monotonic bullion-inflation process. The pre-1600 regime is close to the classical monetary story: money and prices rose together. The post-1600 regime is different: money continued to expand while CPI growth flattened. The appropriate theoretical object is therefore not money quantity alone, but the money-price transmission coefficient.

We begin with the classical one-phase log-level monetary relation:

$$\ln P_t = a + \beta \ln M_t + \varepsilon_t,$$

where $P_t$ is the price level, $M_t$ is money supply, and $\beta$ is the money-price transmission coefficient. This relation can describe a regime in which monetary expansion is transmitted into prices with a stable elasticity. In Spain, it provides a good description of the pre-1600 regime.

The crucial test is whether this one-phase relation remains valid after the transmission break. We therefore estimate the classical relation on the high-transmission phase, 1500–1600, and extrapolate it forward. The resulting gap is

$$G_t = \ln P_t - \ln P_t^{\text{classical}},$$

where $\ln P_t^{\text{classical}}$ is the price level predicted by the pre-1600 one-phase model. If the classical model remains valid, $G_t$ should remain small. If the monetary system enters a different transmission phase, $G_t$ will systematically deviate from zero.

MPTT extends the one-phase relation by adding a second-phase correction term after a transition point $\tau$:

$$\ln P_t = a + \beta_1 \ln M_t + \gamma H(t-\tau)(\ln M_t - \ln M_\tau) + \varepsilon_t,$$

where

$$H(t-\tau) = \begin{cases} 0, & t \le \tau, \\ 1, & t > \tau. \end{cases}$$

Equivalently, the model can be written in hinge form:

$$\ln P_t = a + \beta_1 \ln M_t + \gamma \max(0, \ln M_t - \ln M_\tau) + \varepsilon_t.$$

Before the transition, the correction term is zero and the model reduces to the classical monetary relation:

$$\ln P_t = a + \beta_1 \ln M_t + \varepsilon_t.$$

After the transition, the effective money-price transmission coefficient becomes

$$\beta_2 = \beta_1 + \gamma.$$

Thus, $\gamma$ is the MPTT phase-correction parameter. If $\gamma = 0$, there is no phase correction and the classical one-phase model is sufficient. If $\gamma < 0$, the second phase weakens monetary transmission. If $\gamma > 0$, the second phase amplifies monetary transmission. This is the simplest empirical form of MPTT: the classical monetary model is recovered as the first-phase limit, while the second phase modifies the transmission coefficient.

For Spain, $\tau = 1600$ is used as the historically and visually motivated transition point corresponding to the shift from the high-transmission inflationary phase to the post-1600 decoupling phase. We also conduct an unrestricted break scan over candidate transition years to test whether the data prefer a different break point. Model performance is evaluated using residual error, $R^2$, AIC, and BIC.

# 4. Results

## 4.1 The pre-1600 classical relation works, but fails by extrapolation

Figure 1 and Table 1 establish the two-regime structure of the Spanish Price Revolution. During 1500–1600, CPI increased approximately 3.35-fold, while money supply increased approximately 3.73-fold. The log-level transmission ratio was approximately 0.92, and a direct log-level regression gives a strong positive relation between money and prices. This is the high-transmission metallic inflationary phase.

Table 1. Regime comparison before and after the Spanish monetary-transmission break.

| **Regime** | **Period** | **CPI multiple** | **Money supply multiple** | **Log level transmission ratio lnCPI/lnM** | **Interpretation** |
|---|---|---|---|---|---|
| **I** | 1500–1600 | 3.348 | 3.733 | 0.917 | Classical high-transmission metallic inflationary phase |
| **II** | 1600–1650 | 1.221 | 1.823 | 0.333 | Post-1600 decoupling: money continues to rise while CPI flattens |

However, the classical one-phase relation fails when extended beyond the first regime. Fitting

$$\ln P_t = a + \beta \ln M_t + \varepsilon_t$$

on 1500–1600 gives

$$\beta \approx 0.83.$$

When this pre-1600 relation is extrapolated to 1700, it predicts that CPI should continue rising with money supply (Fig. 2). The observed post-1600 CPI does not follow this extrapolation. The resulting gap becomes negative and persistent after 1600. This is the empirical failure that MPTT is designed to capture.

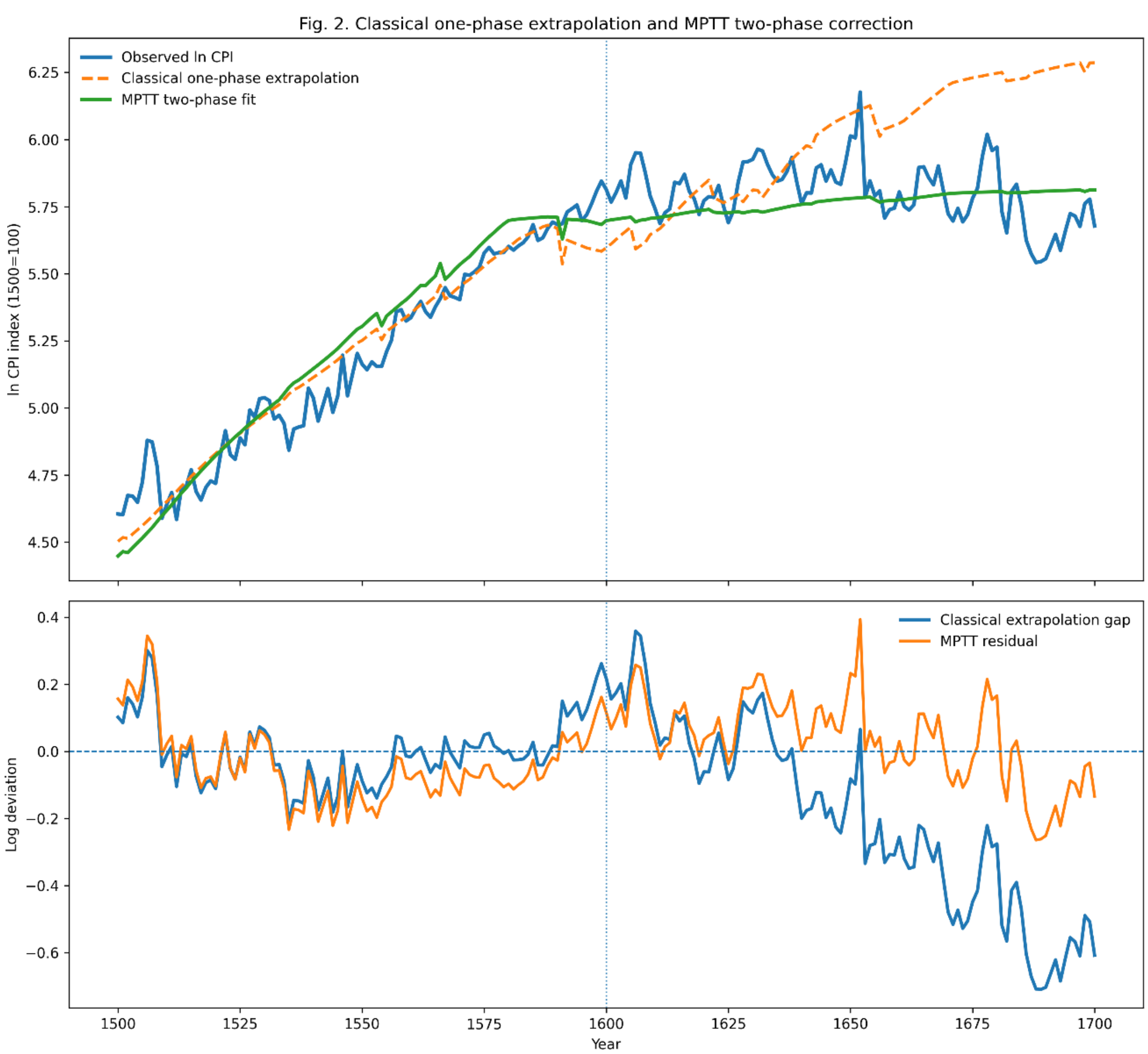

**Fig. 2. Classical one-phase extrapolation and MPTT two-phase correction.**
(A) Observed log CPI, the classical one-phase monetary relation fitted on 1500–1600 and extrapolated to 1700, and the MPTT two-phase fit with $\tau$=1600. The classical relation captures the first regime but overpredicts the price level after 1600. The MPTT correction term captures the post-1600 divergence by allowing the money-price transmission coefficient to fall from $\beta_1 \approx 0.949$ to $\beta_2 \approx 0.137$. (B) The classical extrapolation gap, defined as observed log CPI minus classical predicted log CPI, and the residual from the MPTT two-phase model. The negative post-1600 gap indicates that the classical one-phase model fails by extrapolation, whereas the MPTT residual remains closer to zero.

## 4.2 The MPTT second-phase correction captures the post-1600 gap

The MPTT two-phase model with $\tau = 1600$ fits

$$\ln P_t = a + \beta_1 \ln M_t + \gamma H(t - 1600)(\ln M_t - \ln M_{1600}) + \varepsilon_t.$$

The fitted coefficients are:

$$\beta_1 \approx 0.949,$$

$$\gamma \approx -0.812,$$

and therefore

$$\beta_2 = \beta_1 + \gamma \approx 0.137.$$

The interpretation is direct. Before 1600, the money-price transmission coefficient is close to unity. After 1600, the effective transmission coefficient falls to approximately 0.14. The second-phase correction term is strongly negative, indicating transmission exhaustion rather than transmission amplification.

Figure 2 compares the observed CPI, the classical one-phase extrapolation, and the MPTT two-phase fit. The classical model tracks the first regime but overpredicts prices after 1600. The MPTT correction term absorbs this post-1600 gap and produces a much closer fit to the full 1500–1700 trajectory.

## 4.3 Model comparison supports the phase-correction term

The homogeneous full-sample one-phase model estimates a single coefficient for 1500–1700. It achieves a high overall fit because money and prices still share a broad long-run trend, but it averages across structurally different regimes. The two-phase MPTT model explicitly allows the transmission coefficient to change after the transition.

For the historically motivated $\tau = 1600$ model, the phase correction is large and negative:

$$\gamma \approx -0.812.$$

The model substantially improves fit relative to the full-sample homogeneous relation. This improvement is not merely a matter of adding flexibility: the correction term has a clear interpretation and directly targets the post-1600 divergence between continued monetary expansion and flattened CPI.

## 4.4 Unrestricted break scan identifies a later deep break

We also allow the transition point $\tau$ to vary over candidate years. The BIC-minimizing break is

$$\tau = 1636.$$

At this break, the fitted coefficients are approximately:

$$\beta_1 \approx 0.869,$$

$$\gamma \approx -1.284,$$

so that

$$\beta_2 \approx -0.416.$$

This result does not invalidate the 1600 transition used in the main two-regime interpretation. Rather, it suggests a staged decline in transmission. Around 1600, the high-transmission inflationary phase begins to weaken; by the 1630s, the relationship undergoes a deeper statistical break, consistent with the broader seventeenth-century crisis of the Spanish monarchy. Thus, the empirical pattern is best summarized as:

1. **1500–1600:** high-transmission metallic inflationary phase;
2. **1600–1630s:** weakened transmission / decoupling phase;
3. **after the 1630s:** deeper breakdown or reversal of the earlier money-price relation.

The paper therefore uses 1600 as the historically transparent phase boundary in the main specification, while reporting the unrestricted break scan as evidence that the decline in transmission continues into a deeper seventeenth-century break.

### 4.5 Summary of the new MPTT result

The core result is not that money was irrelevant after 1600. Nor is it that the Spanish Price Revolution was not monetary. The result is that the money-price transmission coefficient changed sharply. The classical monetary relation describes the first phase. It fails by extrapolation after the break. MPTT succeeds by adding a second-phase correction term:

$$\gamma H(t-\tau)(\ln M_t - \ln M_\tau).$$

For Spain, $\gamma < 0$, meaning that the second phase weakens monetary transmission. This is the empirical signature of metallic-money inflation exhaustion.

## 5. Discussion

The Spanish Price Revolution is usually treated as a textbook case of bullion-driven inflation. The results here do not deny that interpretation; they specify its domain of validity. The classical monetary relation works well in the first phase, when money and prices rose together. It fails when extrapolated beyond the transmission break. Spain therefore shows both the validity and the limit of the classical bullion-inflation story.

The MPTT contribution is to make this limit explicit. The phase-correction parameter $\gamma$ measures whether the second phase changes monetary transmissibility. In the Spanish case, $\gamma$ is strongly negative. This means that post-1600 monetary expansion did not disappear, but its price effect was sharply reduced. The relevant historical process is not the absence of money, but the exhaustion of high-transmission metallic-money inflation.

This formulation also clarifies how MPTT differs from earlier SCR theory. SCR emphasizes the functional logic of currency, reservation, and nominal-value preservation. MPTT instead focuses on the empirical transmission coefficient linking money to prices. The two frameworks can be compatible, but they are not identical. SCR is a theory of monetary function; MPTT is a theory of phase-dependent monetary transmission. The Spain paper establishes the latter in a historical metallic-money setting.

The negative phase correction in Spain is historically plausible. The post-1600 weakening occurred after decades of imperial expansion, fiscal stress, war finance, and demographic disruption. Repeated Habsburg bankruptcies, the Dutch Revolt, the Armada defeat, the 1596 crisis cluster, and the plague wave of 1596–1602 all point to an environment in which additional metallic money could be absorbed, diverted, leaked outward, or prevented from producing the same domestic price response as in the earlier high-transmission phase.

Several limitations remain. The transition point $\tau = 1600$ is historically transparent and visually motivated, but the unrestricted BIC scan identifies a later deeper break around 1636. This suggests that transmission exhaustion may have been a staged process rather than a single-year event. The model is deliberately parsimonious and does not include all real-side shocks, fiscal variables, war expenditures, demographic changes, or trade-flow channels. Future work should incorporate these variables explicitly and test whether the phase-correction parameter remains stable under richer specifications.

The broader implication is that the classical monetary relation should be understood as a regime-specific approximation. In one phase, money-price transmission may be strong; in another, it may be weak or even reversed. The MPTT two-phase equation provides a compact empirical language for this phenomenon. It can be applied not only to Spain, but also to modern reserve-dominant expansions, hard-money reservation episodes, and other historical cases where money and prices decouple after a structural transition.

## 6. Conclusion

The Spanish Price Revolution was not a single monotonic bullion-inflation process. It contained a high-transmission metallic inflationary phase before 1600, followed by a weakened-transmission phase after 1600. The classical monetary relation captures the first phase but fails by extrapolation.

The MPTT two-phase equation resolves this failure by adding a second-phase correction term. For Spain, the estimated correction is strongly negative: with $\tau = 1600$, the transmission coefficient falls from approximately 0.949 to approximately 0.137. An unrestricted break scan further suggests a deeper statistical break around 1636.

The conclusion is not that bullion did not matter. It did. The conclusion is that bullion was inflationary only within a specific transmission regime. MPTT formalizes this by treating the classical monetary equation as a first-phase limit and measuring the second-phase correction through $\gamma$. In Spain, $\gamma < 0$, marking the exhaustion of high-transmission metallic-money inflation.

## Data Availability

The data used in this study are derived from publicly available historical datasets and processed supplementary files assembled for this manuscript. The consumer price index series is based on Álvarez-Nogal and Prados de la Escosura (2013). The Spanish money-supply series is based on Chen, Palma, and Ward (2021) and its associated replication materials. The processed datasets generated for the present analysis, including the merged Spain panel, the two-phase core fit data, the model summary table, and the break-scan table, are provided as supplementary data files.

Key processed files include:

- spain_case_core_1492_1810.csv
- spain_mptt_main_table1_two_regime_summary.csv
- spain_mptt_twophase_core_fit_data.csv
- spain_mptt_twophase_model_summary.csv
- spain_mptt_twophase_break_scan.csv
- spain_mptt_historical_timeline_1500_1700.csv

## Acknowledgment

This work is financially supported by the YRI-FD Industrial Project (YRI-IP-25-01).